\documentclass[showpacs,aps,twocolumn]{revtex4}

\usepackage{bm}
\usepackage{amsmath}
\usepackage{graphicx}
\usepackage{subfigure}
\usepackage[usenames,dvipsnames]{color}
\definecolor{darkblue}{RGB}{0,0,196}
\usepackage[colorlinks=true,linkcolor=darkblue,citecolor=darkblue,urlcolor=darkblue]{hyperref}
\usepackage{amsmath,bbm}
\usepackage{amssymb,bm}

\usepackage{hyperref}
\usepackage{footmisc}
\usepackage[makeroom]{cancel}
\usepackage{comment}
\usepackage{lineno}

\usepackage{epsfig}
\usepackage{array}
\usepackage{url}
\usepackage{float}

\definecolor{kugray5}{RGB}{224,224,224}
\usepackage{adjustbox}
\usepackage{multirow}
\definecolor{darkblue}{RGB}{0,0,196}

\newcommand{\bef}{\begin{figure}}
\newcommand{\eef}{\end{figure}}
\newcommand{\bc}{\begin{center}}
\newcommand{\ec}{\end{center}}

\newcommand{\be}{\begin{equation}}
\newcommand{\ee}{\end{equation}}
\newcommand{\bea}{\begin{eqnarray}}
\newcommand{\eea}{\end{eqnarray}}

\def\ba{\begin{eqnarray}}
\def\ea{\end{eqnarray}}

\begin{document}
\title{ Radial Flow and Differential Freeze-out in Proton-Proton Collisions at $\sqrt{s}= 7$ TeV at the LHC}
\author{Arvind Khuntia}
\author{Himanshu Sharma}
\author{Swatantra Kumar Tiwari}
\author{Raghunath Sahoo\footnote{Corresponding author: $Raghunath.Sahoo@cern.ch$}}
%\email{Raghunath.Sahoo@cern.ch}
\affiliation{Discipline of Physics, School of Basic Sciences,\\ Indian Institute of Technology Indore, Indore-453552, India.}
\author{Jean Cleymans}
\affiliation{UCT-CERN Research Centre and Department of Physics, University of Cape Town, Rondebosch 7701, South Africa}

\begin{abstract}
 We analyse the transverse momentum ($p_{\rm T}$)-spectra as a function of charged-particle multiplicity at midrapidity ($|y| < 0.5$) for various identified particles such as $\pi^{\pm}$, $K^{\pm}$, $K_S^0$, $p+\overline{p}$, $\phi$, $K^{*0} + \overline {K^{*0}}$, and $\Lambda$ + $\bar{\Lambda}$ in proton-proton collisions at $\sqrt{s}$ = 7 TeV using Boltzmann-Gibbs Blast Wave (BGBW) model and thermodynamically consistent Tsallis distribution function. We obtain the multiplicity dependent kinetic freeze-out temperature ($T_{\rm kin}$) and radial flow ($\beta$) of various particles after fitting the $p_{\rm T}$-distribution with BGBW model. Here, $T_{\rm kin}$ exhibits mild dependence on multiplicity class while $\beta$ shows almost independent behaviour. The information regarding Tsallis temperature and the non-extensivity parameter ($q$) are drawn by fitting the $p_{\rm T}$-spectra with Tsallis distribution function. The extracted parameters of these particles are studied as a function of charged particle multiplicity density ($dN_{ch}/d\eta$). In addition to this, we also study these parameters as a function of particle mass to observe any possible mass ordering. All the identified hadrons show a mass ordering in temperature, non-extensive parameter and also a strong dependence on multiplicity classes, except the lighter particles. It is observed that as the particle multiplicity increases, the $q$-parameter approaches to Boltzmann-Gibbs value, hence a conclusion can be drawn that system tends to thermal equilibrium. The observations are consistent with a differential freeze-out scenario of the produced particles. 
\end{abstract}
\date{\today}
\maketitle

\section{Introduction}
\label{intro}
High multiplicity proton-proton ($pp$)-collisions at LHC give us the opportunity to study matter under extreme conditions {\it i.e.} at high temperature and/or energy density. Initial energy density results in high pressure gradient, which leads to expansion of the fireball. The interactions among the produced particles are both elastic as well as inelastic, further, depend upon the mean free path of these particles. Recent results on suppression of  $\rm{K}^{*0}/K$ ratio as a function of charged particle multiplicity in $pp$-collisions signifies a presence of hadronic phase in high multiplicity $pp$-collisions with non-zero lifetime \cite{Tripathy:2018ehz}. This hadronic phase is defined as the phase between the chemical freeze-out and kinetic freeze-out. The freeze-out hypersurface, where the inelastic process ceases, known as chemical freeze-out. After chemical freeze-out, elastic collisions are continued till the kinetic freeze-out, where the mean free path of the particles are larger than the system size. This kinetic freeze-out hypersurface can be determined by studying the transverse momentum spectra ($p_{\rm T}$) of the produced particles. The freeze-out processes are complicated and show a hierarchy, where formation of different types of particles and reactions cease at different time scales. From the kinetic theory perspective, reactions with lower interaction cross section switch off early compared to reaction with higher interaction cross section. So, strange and multi-strange particles should freeze-out early as compared to the light flavored hadrons, which leads to a differential freeze-out hypersurfaces. 

Higher probability of multi-partonic interactions at higher collision energies lead to multiple interactions in the produced system, which results in high-multiplicity $pp$-collisions. This multi-partonic interactions might lead to thermalisation in high-multiplicity $pp$-collisions, which can be described by a statistical model.
 
The transverse momentum $(p_{\rm{T}})$ spectra of produced secondaries in high-energy collisions have been proposed to follow a thermalised Boltzmann type of distribution given as~\cite{Hagedorn:1965st},

\begin{eqnarray}
\label{eq1}
E\frac{d^3\sigma}{d^3p}& \simeq C \exp\left(-\frac{p_T}{T_{kin}}\right),
\end{eqnarray}
 where $T_{\rm kin}$ is the kinetic freeze-out temperature.       

The identified particle spectra at RHIC and LHC do not follow Boltzmann-Gibbs distribution due to the possible QCD contributions at high-$p_{\rm T}$. But the low-$p_{\rm T}$-region at high-multiplicity classes in $pp$-collision is explained by incorporating the radial flow ($\beta$) into Boltzmann-Gibbs distribution function, which is known as Boltzmann-Gibbs Blast Wave (BGBW) model~\cite{Schnedermann:1993ws}. The particles in the system are boosted by this radial flow. We can extract $T_{\rm kin}$ and radial flow ($\beta$) by fitting the identified transverse momentum spectra at low-$p_{\rm T}$.

To describe the complete transverse spectra of identified particles, one has to account for the power-law contribution at high-$p_{\rm T}$  \cite{CM,CM1,UA1} and this empirically takes care for the possible QCD contributions.

A combination of both of these aspects has been proposed by Hagedorn, which describes the experimental data over a wide  $p_{\rm T}$-range  \cite{Hagedorn:1983wk} and is given by 
  
\begin{eqnarray}
  E\frac{d^3\sigma}{d^3p}& = &C\left( 1 + \frac{p_T}{p_0}\right)^{-n}
\nonumber\\
 & \longrightarrow&
  \left\{
 \begin{array}{l}
  \exp\left(-\frac{n p_T}{p_0}\right)\quad \, \, \, {\rm for}\ p_{\rm T} \to 0, \smallskip\\
  \left(\frac{p_0}{p_T}\right)^{n}\qquad \qquad{\rm for}\ p_{\rm T} \to \infty,
 \end{array}
 \right .
 \label{eq2}
\end{eqnarray}
where $C$, $p_0$, and $n$ are fitting parameters. 

The resultant expression acts as an exponential and a power-law function for small and large $p_{\rm T}$, respectively. A deviation is observed by experiments at RHIC~\cite{star-prc75,phenix-prc83} and LHC~\cite{alice1,alice2,alice3,cms} while describing the $p_{\rm T}$-spectra of identified particles using the equilibrium statistical distribution function. The mean transverse momentum ($\langle p_{T} \rangle$) of the equilibrated hadronic matter is related with the temperature but one cannot establish this relation for the systems away from thermal equilibrium. In case of the latter systems, the temperature fluctuates either event-by-event or within the same event~\cite{Bhattacharyya:2015nwa}. This necessitates to use non-extensive Tsallis statistics for the description of the $p_{\rm T}$-spectra in high-energy hadronic and nuclear collisions~\cite{Tsallis:1987eu,Tsallis:2008mc,book}. A thermodynamically consistent Tsallis non-extensive distribution function at mid-rapidity is given as follows~\cite{Cleymans:2011in}, 

\begin{equation}
\label{eq3}
f(m_T) =  C_q \left[1+{(q-1)}{\frac{m_T}{T}}\right]^{-\frac{1}{q-1}} .
\end{equation}
 Here, $m_{\rm T} = \sqrt{p_T^2 + m^2}$ is the transverse mass and $q$ is the non-extensive parameter, which measures the degree of deviation from equilibrium. Eqs. \ref{eq2} and \ref{eq3} are related via the following equations for the large values of $p_{\rm T}$,
  \begin{equation}
  n= \frac{1}{q-1}, ~\mathrm{and} ~~~~ p_0 = \frac{T}{q-1}.
  \label{eq4}
  \end{equation} 
  
 In the limit of $q \rightarrow 1$, Tsallis distribution (Eq.~\ref{eq3}) reduces to the standard Boltzmann-Gibbs distribution (Eq.~\ref{eq1}). 
      
Tsallis distribution is widely used to explain the particle spectra of produced particles in high-energy collisions~\cite{Bhattacharyya:2015nwa,Bhattacharyya:2015hya,Zheng:2015gaa,Tang:2008ud,De:2014dna} such as in elementary 
$e^++e^-$, hadronic and heavy-ion collisions~\cite{e+e-,R1,R2,R3,ijmpa,plbwilk,marques,STAR,PHENIX1,PHENIX2,ALICE_charged,ALICE_piplus,CMS1,CMS2,ATLAS,ALICE_PbPb}. The criticality in the non-extensive $q$-parameter is also observed while studying the speed of sound for hadron gas using non-extensive statistics~\cite{Khuntia:2016ikm}. Recently, comprehensive studies have been carried out for $\pi^-$ and quarkonium spectra in $pp$-collisions~\cite{Grigoryan:2017gcg,Parvan:2016rln}.
 
The paper is organised as follows. In section~\ref{sec:1},  we present  the BGBW to  describe the identified particle spectra  upto $p_{\rm{T}} ~$$\simeq$ 3 GeV/$c$ and also discuss about $T_{\rm kin}$ and radial flow ($\beta$) as a function of charged particle multiplicity and particle mass. Similarly, in section~\ref{sec:2}, we use a thermodynamically consistent Tsallis distribution function to  describe the identified particle spectra for the complete range of $p_{\rm{T}} $.  Also we discuss the results  in view of the non-extensive statistical description of the identified particles as a function of charged particle multiplicity and particle mass. Finally, in section~\ref{sec3} we present the summary of our results.

%%%%%%%%%%%%%%%%%%%%%%%%%%%%%%%%%%%%%%%%%%%%%%%%%%%%%%%%%%%%%%%%%%%%%%%%%%%%%

%%%%%%%%%%%%%%%%%%%%%%%%%%%%%%%%%%%%%%%%%%%%%%%%%%%%%%%%

\section{Transverse momentum spectra of Identified Hadrons}

In this section, we analyse the transverse momentum spectra of $\pi^{\pm}$, $K^{\pm}$, $K^{0}_{S}$, $p+\overline{p}$, $\phi$, $K^{*0} + \overline {K^{*0}}$, and $\Lambda+\bar{\Lambda}$ produced in $pp$-collisions at $\sqrt {s}$ = 7 TeV at the LHC, measured by the ALICE~\cite{Acharya:2018orn,Adam:2016emw}. This $p_{\rm T}$-spectra is analysed by using Boltzmann-Gibbs Blast Wave (BGBW) model and a thermodynamically consistent Tsallis non-extensive statistics.

%%%%%%%%%%%%%%%%%%%%%%%%%%%%%%%%%%%%%%%%%%%%%%%%%%%%%%%%%

\subsection{Boltzmann-Gibbs Blast Wave (BGBW) Model}
\label{sec:1}

In this subsection, we employ BGBW model to fit the transverse momentum spectra of various identified light flavor hadrons measured at $\sqrt{s}$ = 7 TeV. For this study, we do not include multistrange particles because statistical uncertainties forbids us in drawing any physics conclusion. The expression for invariant yield in the framework of BGBW is given as follows~\cite{Schnedermann:1993ws}:
 
 \ba
\label{bgbw1}
E\frac{d^3N}{dp^3}=D \int d^3\sigma_\mu p^\mu exp(-\frac{p^\mu u_\mu}{T}),
\ea

where the particle four-momentum is, 

\ba
p^\mu~=~(m_T{\cosh}y,~p_T\cos\phi,~ p_T\sin\phi,~ m_T{\sinh}y), 
\ea
the four-velocity is given by,
\ba
u^\mu=\cosh\rho~(\cosh\eta,~\tanh\rho~\cos\phi_r,~\tanh\rho~\sin~\nonumber\\
\phi_r,~\sinh~\eta),
\ea 
while the kinetic freeze-out surface is parametrised as, 

\ba
d^3\sigma_\mu~=~(\cosh\eta,~0,~0, -\sinh\eta)~\tau~r~dr~d\eta~d\phi_r.
\ea

Here, $\eta$ is the space-time rapidity. With simplification assuming Bjorken correlation in rapidity, $i.e.$ $y=\eta$~\cite{Bjorken:1982qr}, Eq.~\ref{bgbw1} can be expressed as:  
\ba
\label{boltz_blast}
\left.\frac{d^2N}{dp_Tdy}\right|_{y=0} = D \int_0^{R_{0}} r\;dr\;K_1\Big(\frac{m_T\;\cosh\rho}{T_{kin}}\Big)I_0\nonumber\\
\Big(\frac{p_T\;\sinh\rho}{T_{kin}}\Big),
\ea
where $D$ is the normalisation constant. Here $g$ is the degeneracy factor and $m_{\rm T}=\sqrt{p_T^2+m^2}$ is the transverse mass. $K_{1}\displaystyle\Big(\frac{m_T\;{\cosh}\rho}{T_{kin}}\Big)$ and $I_0\displaystyle\Big(\frac{p_T\;{\sinh}\rho}{T_{kin}}\Big)$ are the modified Bessel's functions and are given by,

%\begin{widetext}
\ba
\centering
K_1\Big(\frac{m_T\;{\cosh}\rho}{T}\Big)=\int_0^{\infty} {\cosh}y\;{\exp} \Big(-\frac{m_T\;{\cosh}y\;{\cosh}\rho}{T_{kin}}\Big)dy\nonumber,
\ea

\ba
\centering
I_0\Big(\frac{p_T\;{\sinh}\rho}{T}\Big)=\frac{1}{2\pi}\int_0^{2\pi} exp\Big(\frac{p_T\;{\sinh}\rho\;{\cos}\phi}{T_{kin}}\Big)d\phi \nonumber,
\ea
%\end{widetext}
where $\rho$ in the integrand is a parameter given by $\rho={\tanh}^{-1}\beta$, with $\beta=\displaystyle\beta_s\;\Big(\xi\Big)^n$ \cite{Huovinen:2001cy,Schnedermann:1993ws,BraunMunzinger:1994xr, Tang:2011xq} is the radial flow. $\beta_s$ is the maximum surface velocity and $\xi=\displaystyle\Big(r/R_0\Big)$, with $r$ as the radial distance. In the blast-wave model the particles closer to the center of the fireball move slower than the ones at the edges. The average of the transverse velocity can be evaluated as \cite{Adcox:2003nr}, 
\ba
<\beta> =\frac{\int \beta_s\xi^n\xi\;d\xi}{\int \xi\;d\xi}=\Big(\frac{2}{2+n}\Big)\beta_s.
\ea
In our calculation, we use a linear velocity profile, ($n=1$) and $R_0$ is the maximum radius of the expanding source at freeze-out ($0<\xi<1$). 

\bef[H]
\begin{center}
\includegraphics[scale=0.35]{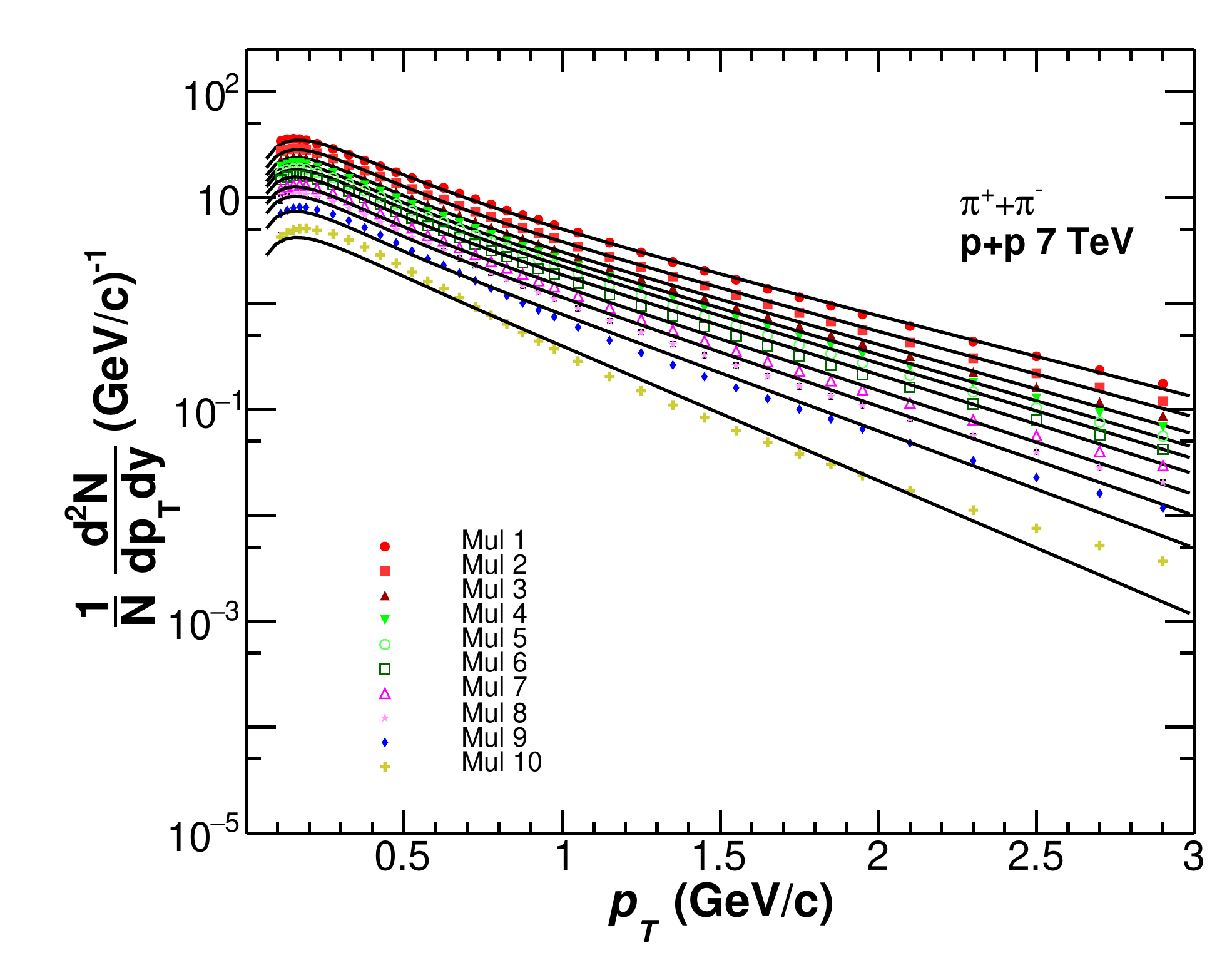}
\includegraphics[scale=0.35]{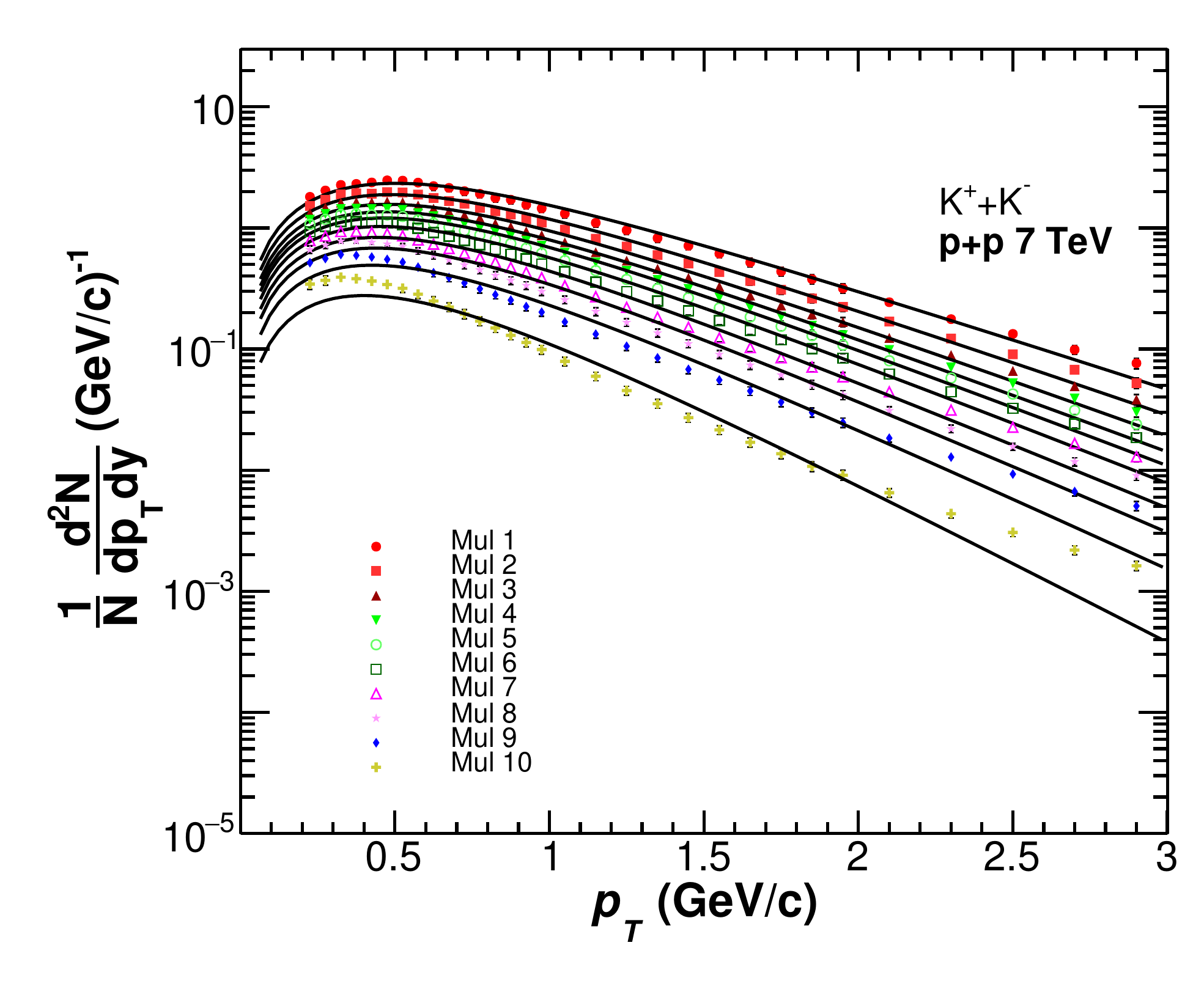}
\includegraphics[scale=0.35]{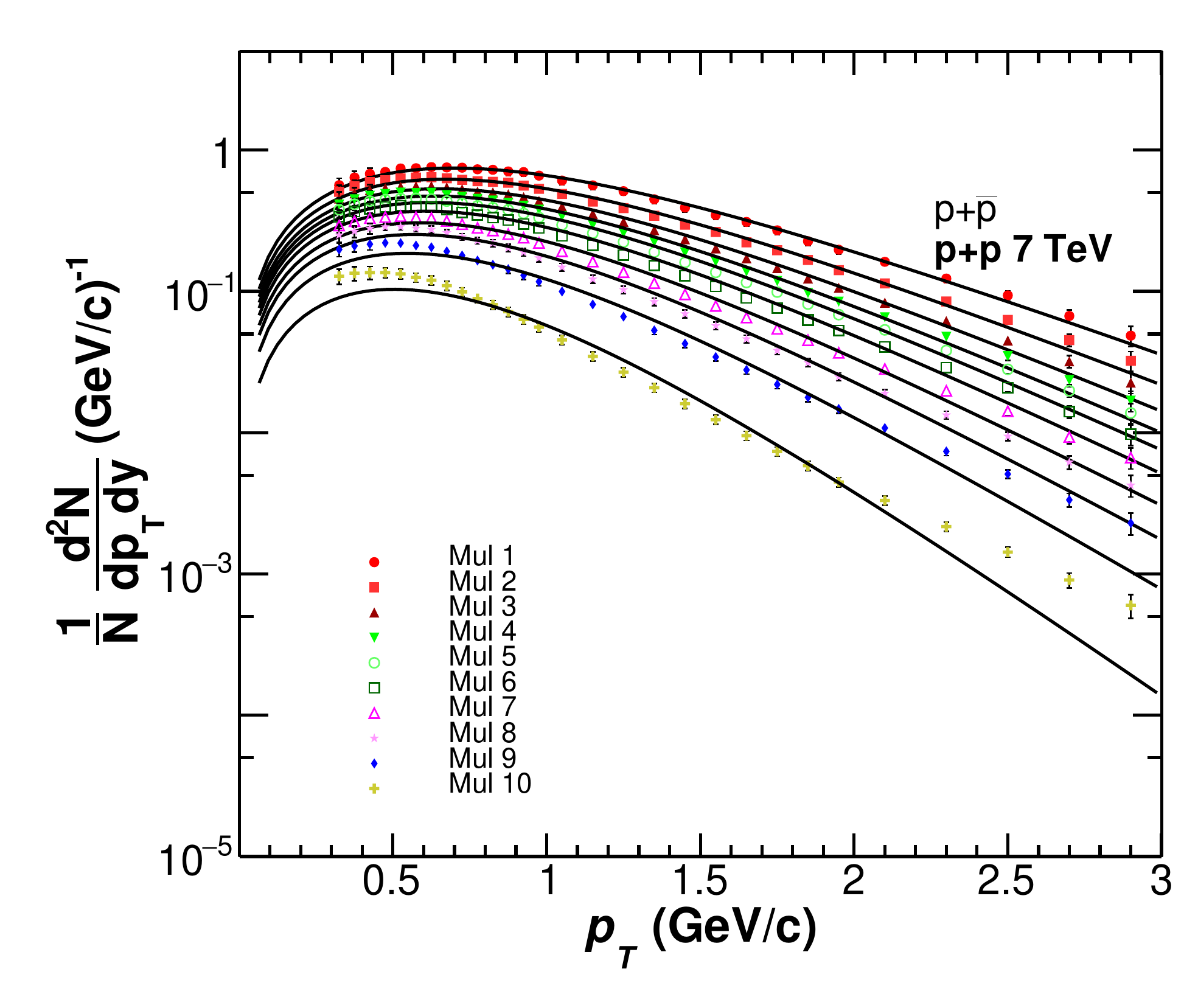}
\caption{(Color online) Fitting of experimentally measured $p_{\rm T}$-spectra of pion ($\pi^{\pm}$), kaon ($K^{\pm}$) and proton ($p+\overline{p}$)~\cite{Acharya:2018orn} with BGBW model for $pp$-collisions at $\sqrt{s}$ = 7 TeV using Eq.~\ref{boltz_blast} for various multiplicity classes, as defined in Table 1.}
\label{BG}
\end{center}
\eef

\bef[ht]
\begin{center}
\includegraphics[scale=0.40]{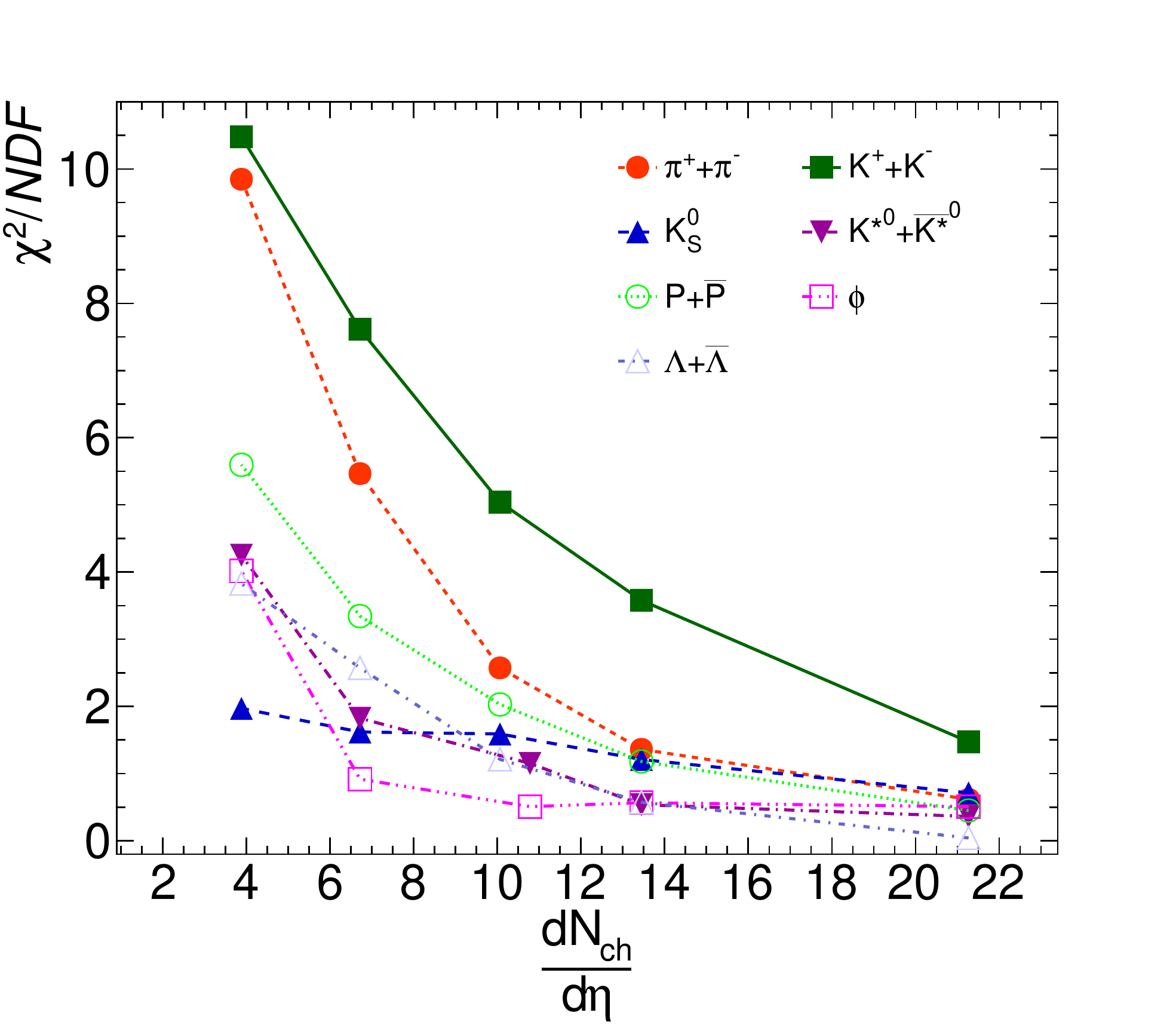}
\newline
\caption{(color online) $\chi^2$/NDF for various identified particles as a function of charged particle multiplicity classes extracted in BGBW model.}
\label{chiBGBW}
\end{center}
\eef

In figure~\ref{BG}, we show the fitting of $p_{\rm T}$-spectra of pion ($\pi^{\pm}$), kaon ($K^{\pm}$) and proton ($p+\overline{p}$)~\cite{Acharya:2018orn} measured in $pp$-collisions at $\sqrt{s}$ = 7 TeV in various multiplicity classes. Here, we have used BGBW function to fit the spectra given by Eq.~\ref{boltz_blast}, where $T_{\rm kin}$ and radial flow ($\beta$) are the free parameters. We observe that BGBW function explains the experimental data reasonably well for all the particles at lower $p_{\rm T}$ ($\sim$ 3 GeV/$c$). In Fig.~\ref{chiBGBW}, we have shown the extracted $\chi^2$/NDF for all the considered particles. We find a good $\chi^2$/NDF for all hadrons except $\pi^{\pm}$ and $K^{\pm}$. This may be due to the fact that resonance decay effects are not taken care of in the BGBW model, which affects $\pi^{\pm}$ and $K^{\pm}$ more compared to higher mass particles like $p$ and $\Lambda$ etc. The fitting of BGBW worsens for all identified particles, when one goes for lower multiplicity classes, as shown explicitly in Fig. \ref{chiBGBW}. This is because of the smaller particle multiplicities available in the lower multiplicity classes. Additionally, we observe deviations of BGBW expectations from the experimental data at higher-$p_{\rm T}$ for lower multiplicity classes. 

\bef[ht]
\begin{center}
\includegraphics[scale=0.40]{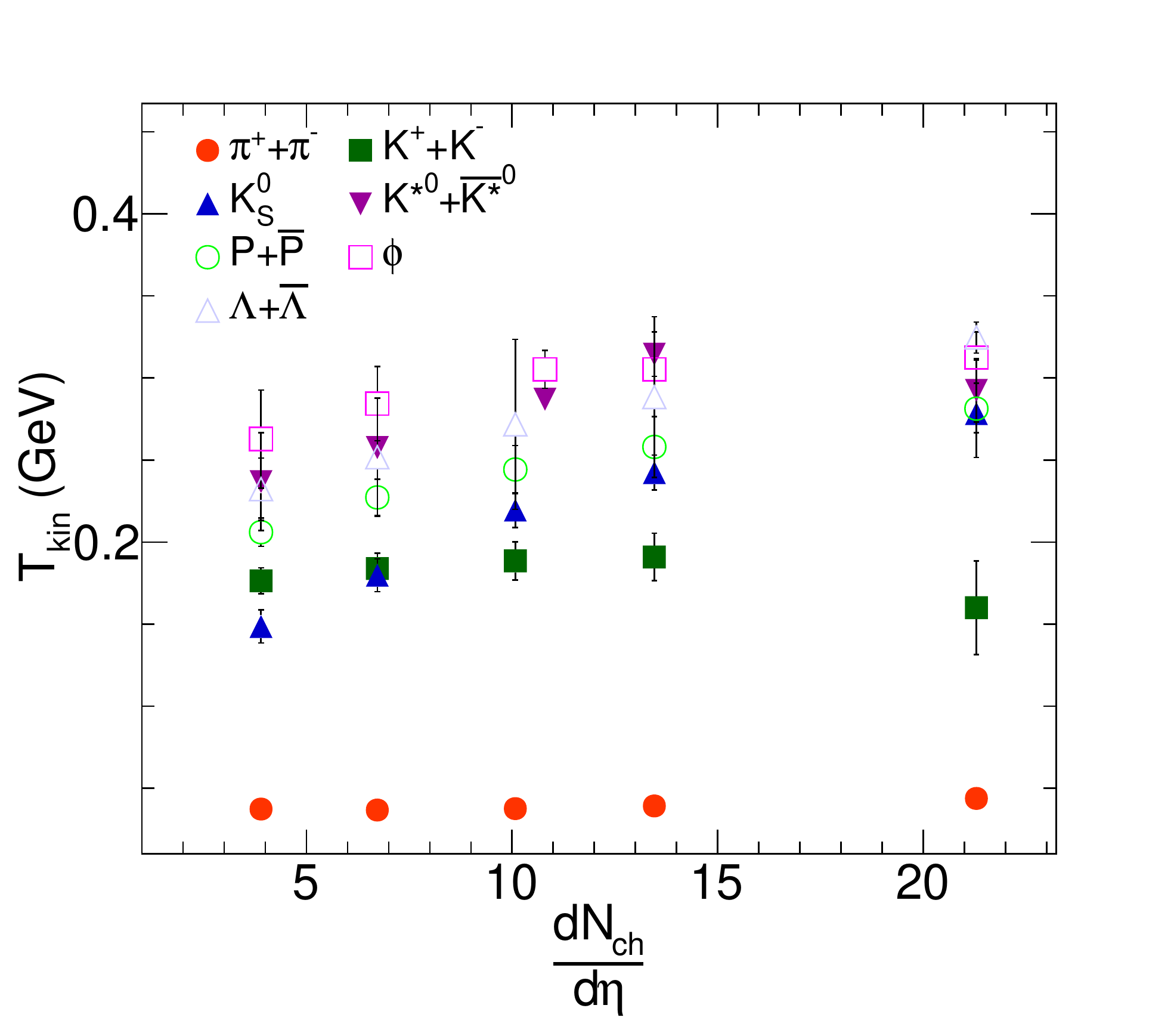}
\caption{(color online) Multiplicity dependence of $T_{\rm kin}$ for $pp$-collisions at $\sqrt{s}$ = 7 TeV using Eq.~\ref{boltz_blast} as the fitting function. }
\label{BGT}
\end{center}
\eef

\bef[ht]
\begin{center}
\includegraphics[scale=0.40]{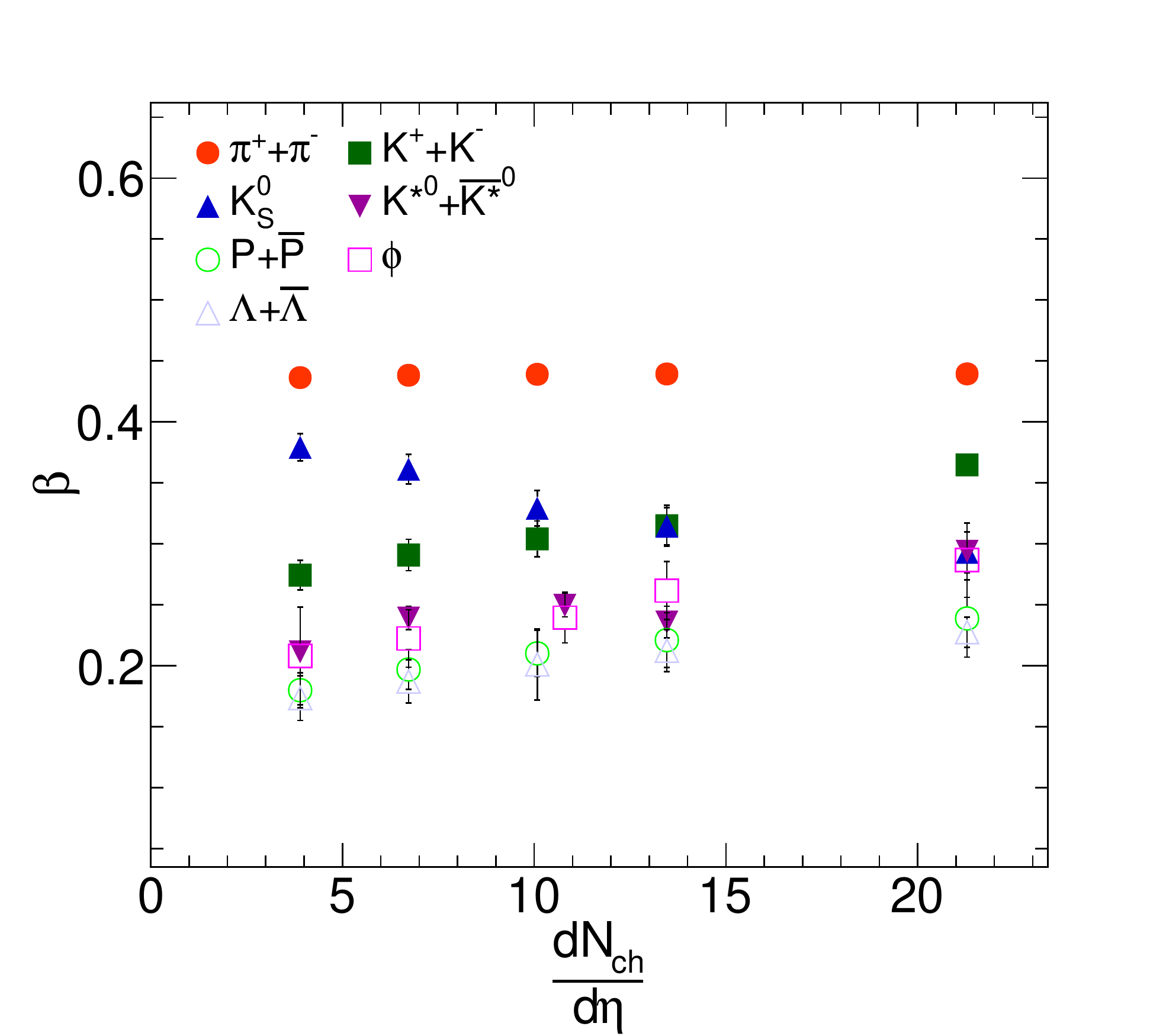}
\newline
\caption{(color online) Multiplicity dependence of the radial flow, $\beta$ for $pp$-collisions at $\sqrt{s}$ = 7 TeV using Eq.~\ref{boltz_blast} as the fitting function.}
\label{BGbeta}
\end{center}
\eef

We have extracted the kinetic freeze-out temperature and radial flow velocity for all the hadrons considered here. Figure~\ref{BGT} represents the extracted $T_{\rm kin}$ as a function of charged particle multiplicity for identified hadrons. We notice that, $T_{\rm kin}$ is higher for massive particles in comparison to lighter ones which supports the differential freeze out scenario and suggest that massive particles freeze-out earlier from the system. However, $K^0_S$ behaves differently in the system as reported earlier \cite{Arvind:2017}.

In figure~\ref{BGbeta}, we have demonstrated the extracted radial flow velocity for the identified particles as a function of multiplicity class. We find that $\beta$ value reduces as the particle mass increases in all the multiplicity classes except for proton. Again, $\rm{K}^0_S$ behaves in a different manner which is not understood in the present work. Further, almost equal magnitude of radial flow is observed for $\rm{K}^{*0}$ and $\phi$, which have similar masses. Furthermore, we observe that $p$ has mass similar to $\rm{K}^{*0}$ and $\rm{\phi}$ but the radial flow of $p$ is closer to $\Lambda$, as both $p$ and $\Lambda$ are baryons. This advocates that baryons freeze-out at a similar temperature, while mesons such as $\phi$ and $K^{*0}$ having nearly the same mass freeze-out from a different temperature. These observations indicate that while respecting the baryon-meson freeze-out, particles obey a hydrodynamic behavior.

\bef[H]
\begin{center}
\includegraphics[scale=0.35]{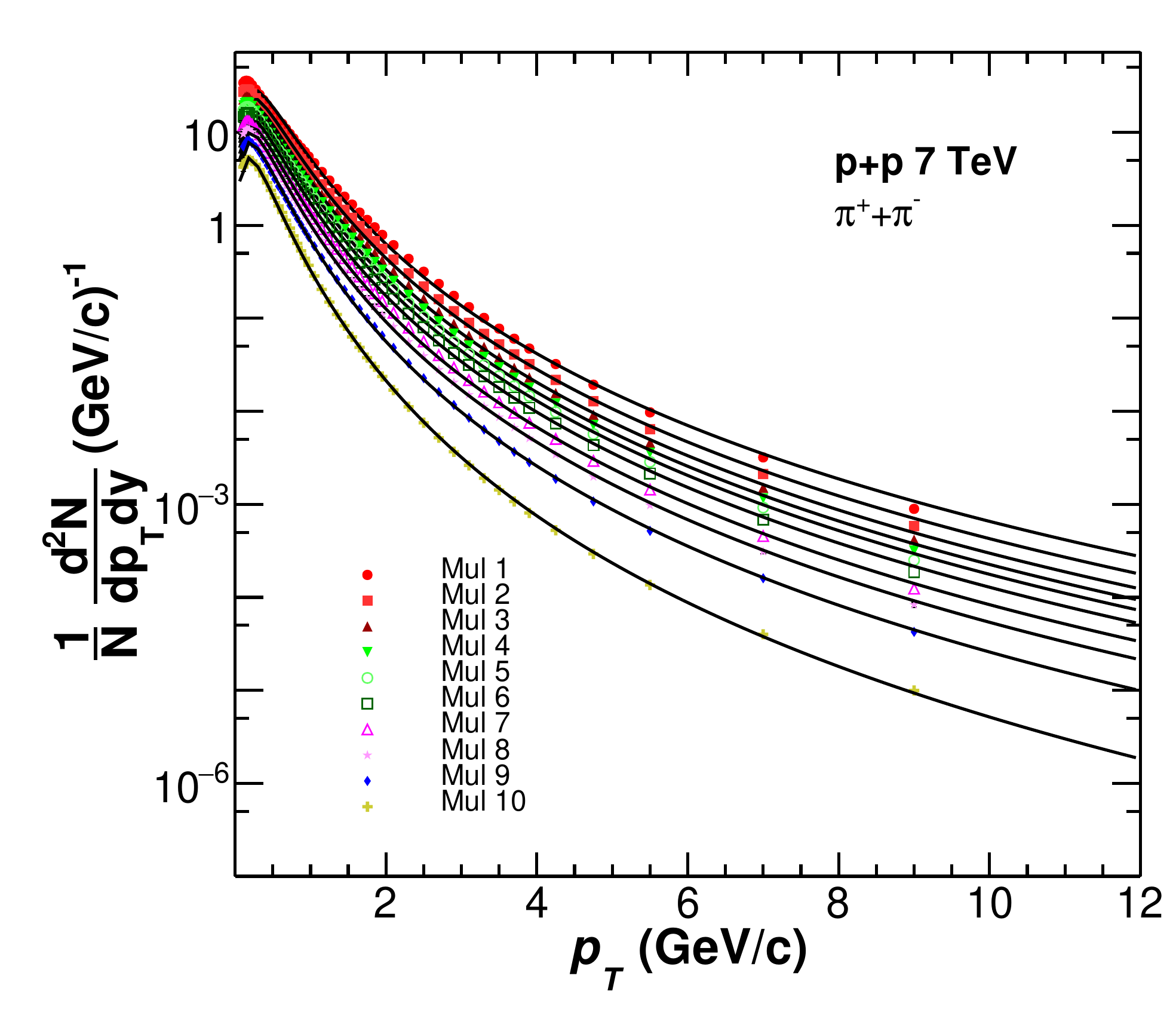}
\includegraphics[scale=0.35]{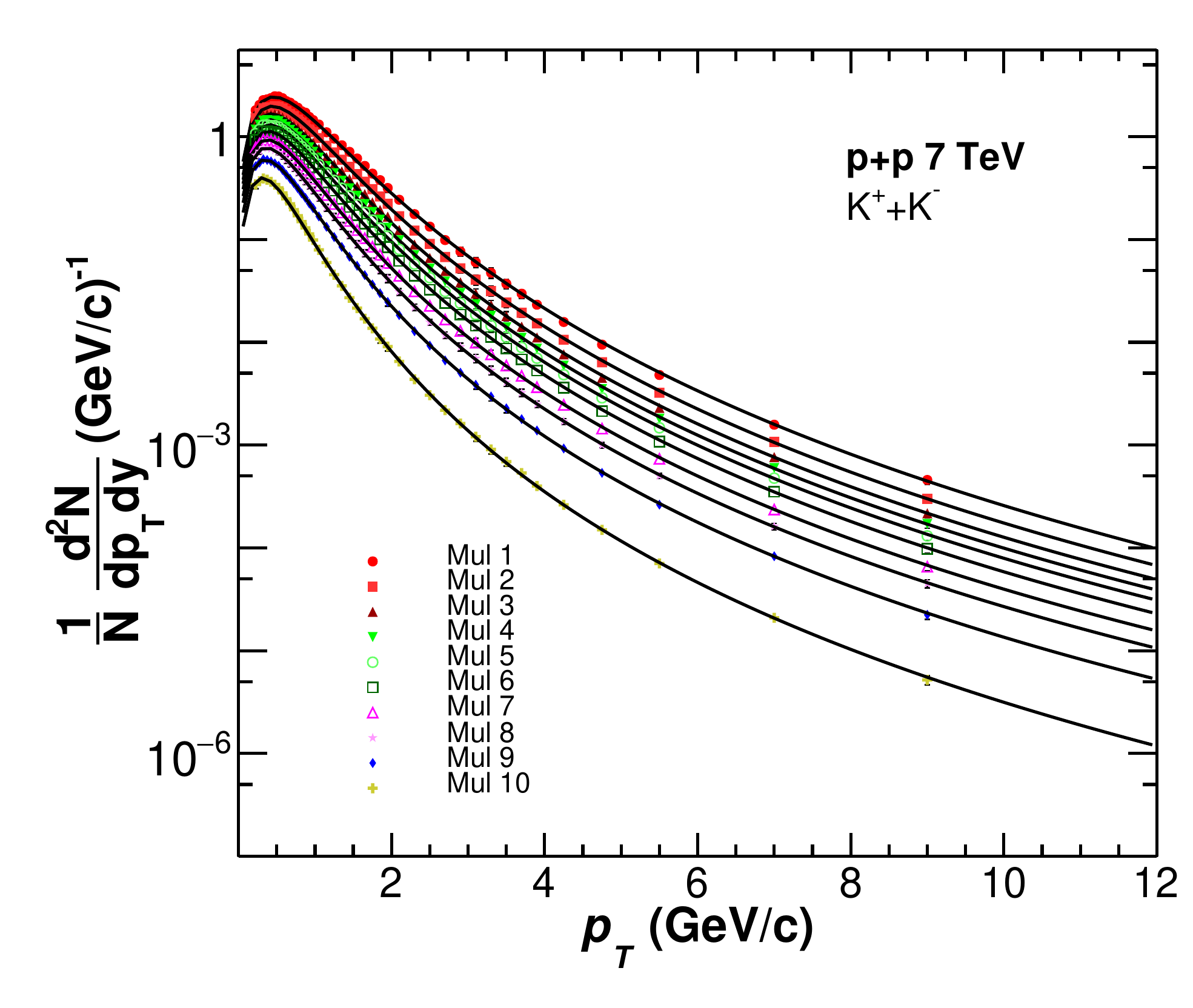}
\includegraphics[scale=0.35]{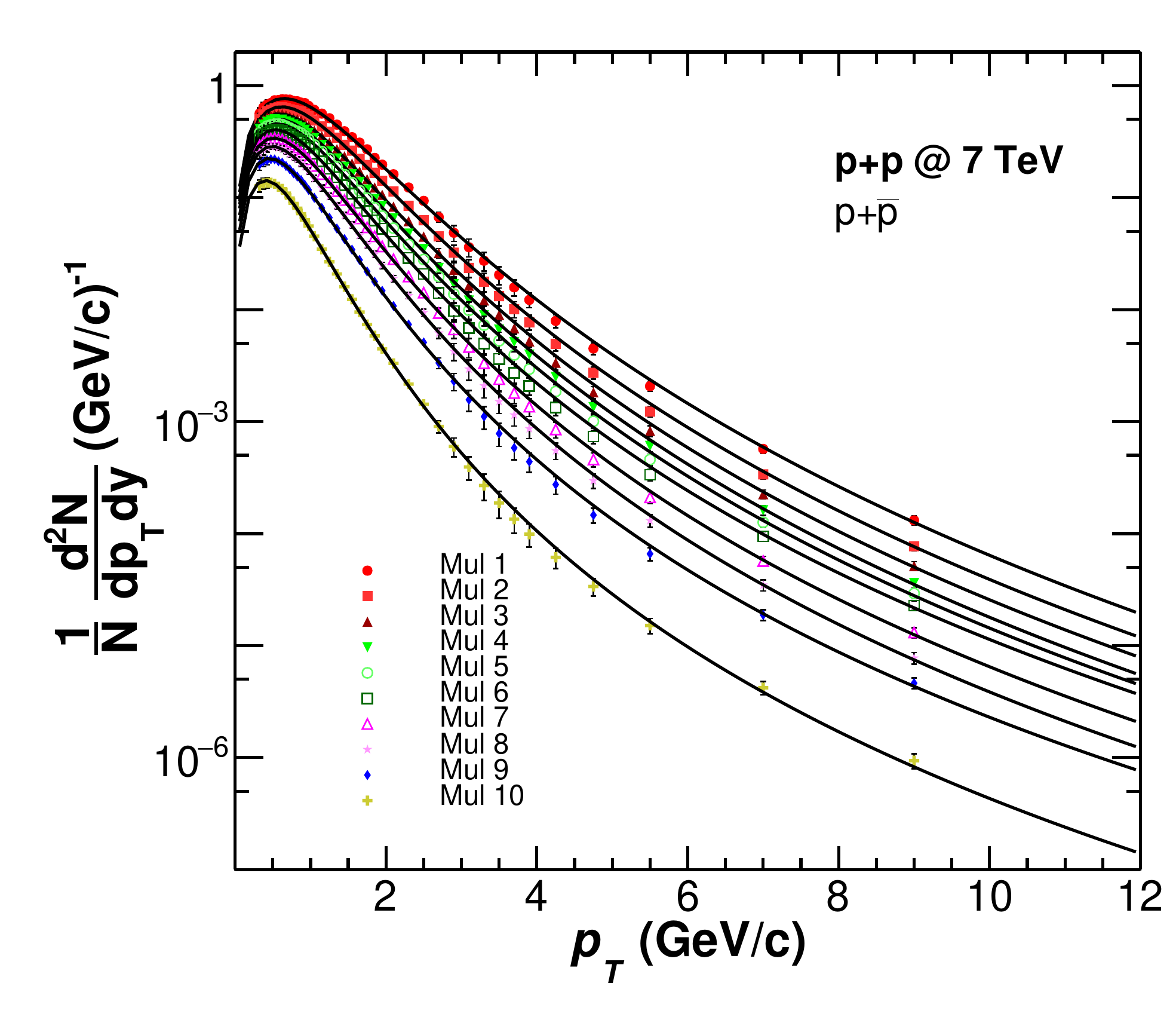}
\caption{(Color online) Experimentally measured $p_{\rm T}$-spectra of pion ($\pi^{\pm}$), kaon ($K^{\pm}$) and proton ($p+\overline{p}$)~\cite{Acharya:2018orn}, fitted with Tsallis distribution for $pp$-collisions at $\sqrt{s}$ = 7 TeV using Eq.~\ref{eq6} for various multiplicity classes. The extracted parameters are given in table~\ref{table:parameters}.}
\label{pikp}
\end{center}
\eef

\bef[H]
\begin{center}
\includegraphics[scale=0.35]{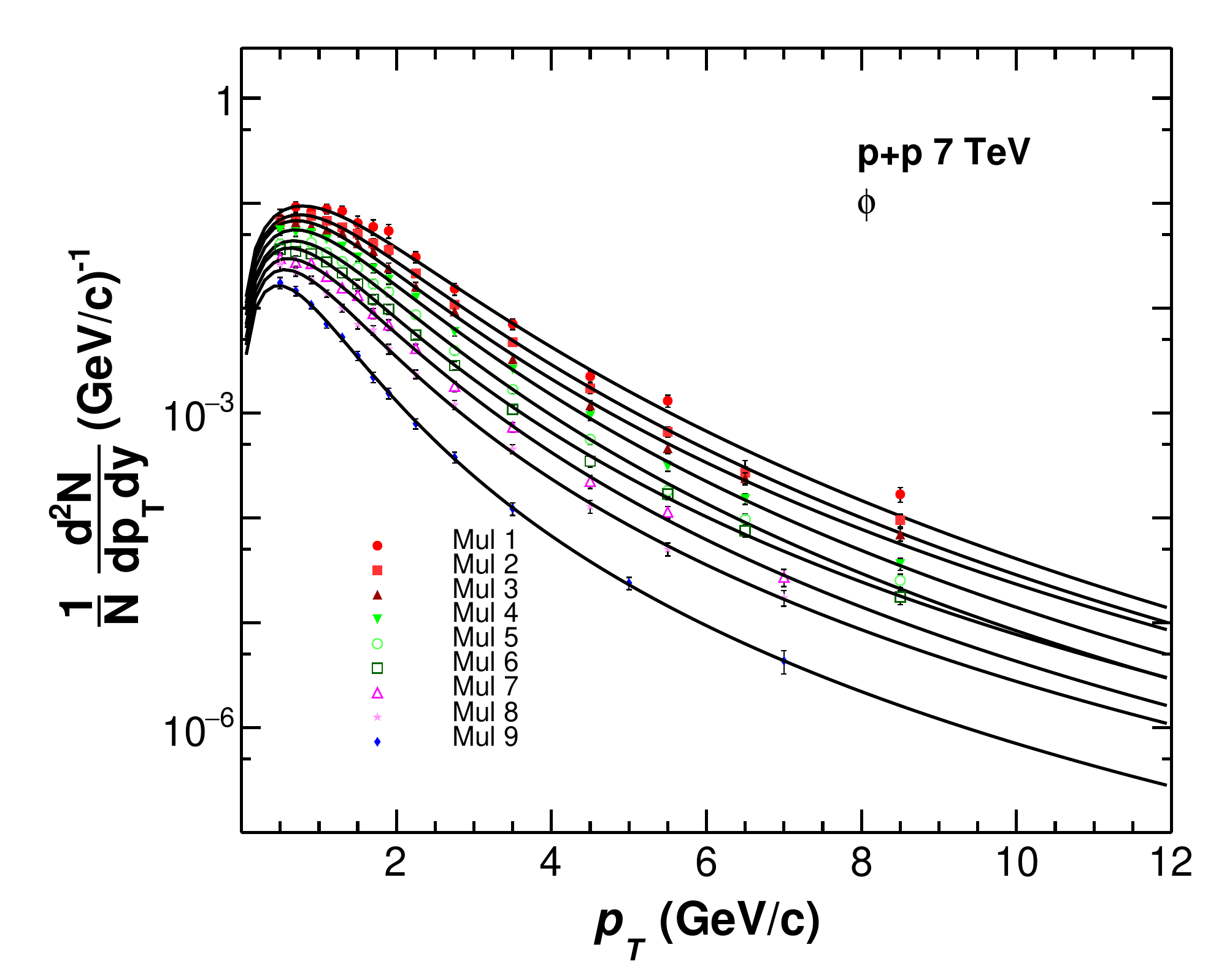}
\includegraphics[scale=0.35]{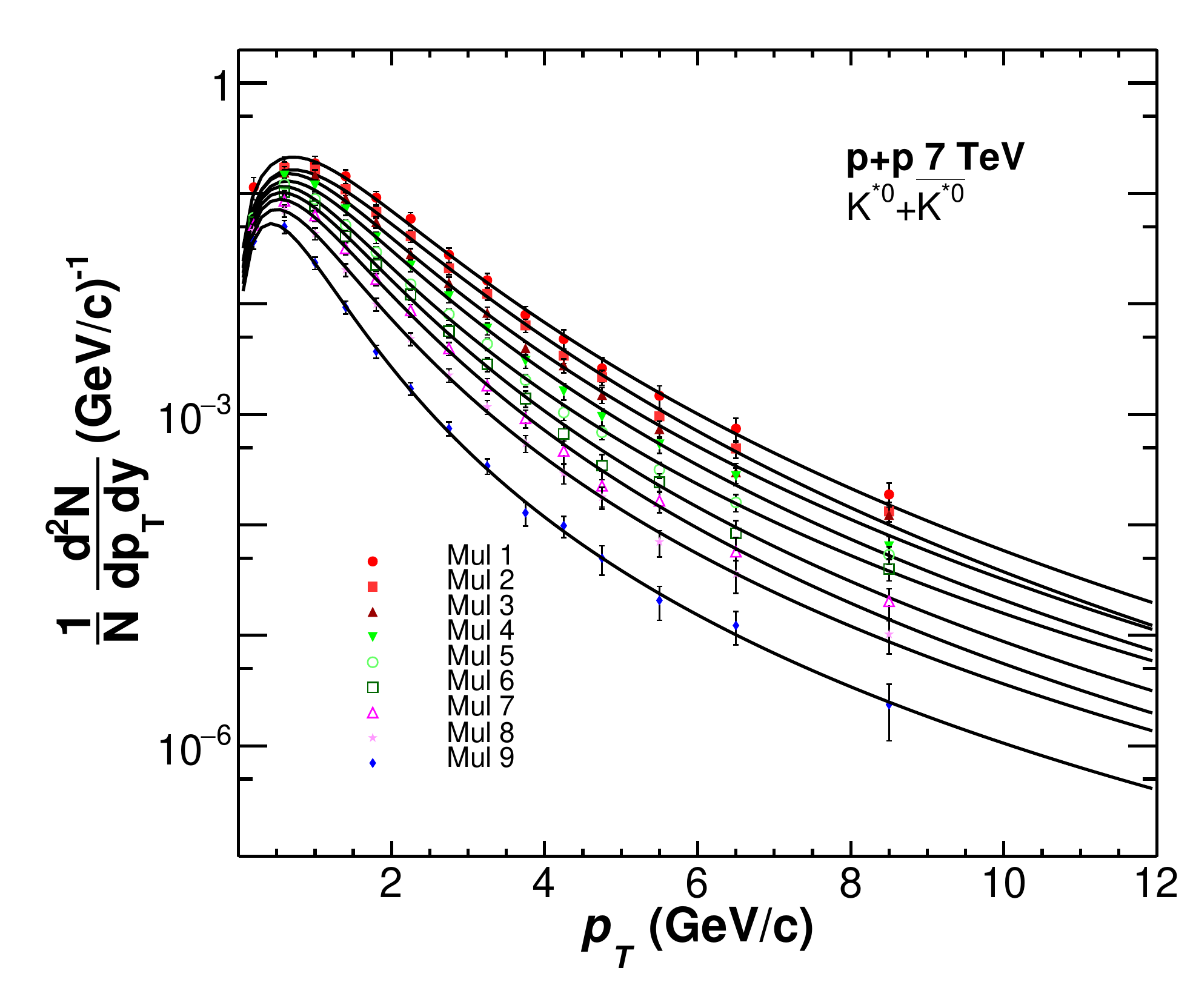}
\caption{(Color online) The $p_{\rm T}$-spectra of $\phi$ and $K^{*0} + \overline {K^{*0}}$ ~\cite{Acharya:2018orn} fitted with Tsallis distribution for $pp$-collisions at $\sqrt{s}$ = 7 TeV using Eq.~\ref{eq6} for various multiplicity classes. The extracted parameters are given in table~\ref{table:parameters}.}
\label{phiKstar}
\end{center}
\eef

%%%%%%%%%%%%%%%%%%%%%%%%%%%%%%%%%%%%%%%%%%%%%%%%%%%%%%%%%%

\subsection{Non-extensivity and $p_{\rm T}$-spectra}
\label{sec:2}
The Tsallis distribution function at mid-rapidity, with finite chemical potential  \cite{Cleymans:2015lxa} is given by,

\begin{eqnarray}
\label{tsal}
\left.\frac{1}{p_T}\frac{d^2N}{dp_Tdy}\right|_{y=0} = \frac{gVm_T}{(2\pi)^2}
\left[1+{(q-1)}{\frac{m_T-\mu}{T}}\right]^{-\frac{q}{q-1}},
\end{eqnarray}
 where, $V$ is the system volume and $\mu$ is the chemical potential of the system. For $\mu$ = 0, the above equation~\ref{tsal} modifies as~\cite{Li:2015jpa},
\begin{eqnarray}
\label{eq6}
\left.\frac{1}{p_T}\frac{d^2N}{dp_Tdy}\right|_{y=0} = \frac{gVm_T}{(2\pi)^2}
\left[1+{(q-1)}{\frac{m_T}{T}}\right]^{-\frac{q}{q-1}}.
\end{eqnarray}
Now, we use Eq.~\ref {eq6} to fit the transverse momentum spectra of various particles measured experimentally for different multiplicity classes at $\sqrt{s}$ = 7 TeV. The fittings are shown in Figs.~\ref{pikp} and \ref{phiKstar}. We have listed the definitions of the multiplicity classes in Table ~\ref{table:mult_info}. The fitting is performed using the TMinuit class available in ROOT library keeping all the parameters free. Here, we follow the notion of a mass dependent differential freeze-out scenario \cite{Thakur:2016boy,Lao:2015zgd}, where particles freeze-out at different times, which correspond to different system volumes and temperatures. Henceforth, we study the thermodynamic parameters in the context of non-extensive statistics. After fitting we found that, $\chi^2$/NDF is below 1 for all the considered particles for highest multiplicity while it increases as the multiplicity decreases, as is tabulated in Table 2. This shows that the spectra are very well described by the non-extensive statistics particularly at highest multiplicity class.

Figure~\ref{fit:Tsallis:T} represents the temperature parameter, $T$ extracted in the fitting as a function of event multiplicity for all the considered particles. We notice a monotonic increase in $T$ with the increase in particle multiplicity for all the hadrons. For heavier particles, the temperature is observed to be higher, which indicates an early freeze-out of these particles. We also find that the temperature for lighter particles does not change appreciably with multiplicity but for heavier particles it shows a significant variation of $T$ with charged particle multiplicity. We have also shown the variation of the non-extensive parameter, $q$ with charged particle multiplicity in Fig.~\ref{fit:Tsallis:q}. The value of $q$ decreases monotonically for higher multiplicity classes for all the particles discussed in this work. These findings suggest that the system formed in higher multiplicity class is close to thermal equilibrium. However for $\pi^{\pm}$, $K^{\pm}$ and $K^{0}_{S}$, $q$ is almost independent of the charged particle multiplicity density. The fact that the $q$-values go on decreasing with multiplicity is an indicative of the tendency of the produced systems towards thermodynamic equilibrium. This goes inline with the naive expectations while understanding the microscopic view of systems approaching thermodynamic equilibrium. A similar tendency of $q$ decreasing with number of participating nucleons for Pb+Pb collisions at $\sqrt{s_{\rm NN}}$ = 2.76 TeV has been observed for the bulk part ($p_T <$ 6 GeV/$c$ ) of the charged hadron spectra \cite{Biro:2014cka,Urmossy:2015hva}. The present study is useful in understanding the microscopic features of degrees of equilibration and their dependencies on the number of particles in the system. 
\begin{table*}[htbp]
\caption[p]{Number of mean charged particle multiplicity density corresponding to different event classes \cite{Adam:2016emw,Acharya:2018orn}.}
\label{table:mult_info}
\begin{adjustbox}{max width=\textwidth}
\begin{tabular}{c|c|c|c|c|c|c|c|c|c|c|c|}
\hline
\multicolumn{2}{|c|}{${\bf Class name}$}&Mul1&Mul2&Mul3&Mul4&Mul5&Mul6&Mul7&Mul8&Mul9&Mul10\\
\hline

\multicolumn{2}{|c|}{ $  \bf \big<{\frac{dN_{ch}}{d\eta} } \big>$} &21.3$\pm$0.6&16.5$\pm$0.5&13.5$\pm$0.4&11.5$\pm$0.3&10.1$\pm$0.3&8.45$\pm$0.25&6.72$\pm$0.21&5.40$\pm$0.17&3.90$\pm$0.14&2.26$\pm$0.12\\
\hline
\end{tabular}
\end{adjustbox}
 \end{table*}
 
\begin{table*}[htbp]
\caption{The extracted Tsallis parameters as well as the $\chi^2/NDF$  for all the multiplicity classes. For $\Omega^-+\bar{\Omega^+}$, multiplicity classes are combined to deal with the statistics~\cite{Adam:2016emw}, For  $K_{S}^{0}$, $\Lambda$ + $\bar{\Lambda}$, $\Xi^{-}+\bar{\Xi}^{+}$ and $\Omega^-+\bar{\Omega^+}$, the parameters are taken from our earlier work~\cite{Arvind:2017}.}
\label{table:parameters}
\newcommand{\tabincell}
\centering
\begin{adjustbox}{max width=\textwidth}
\begin{tabular}{|c|c|c|c|c|c|c|c|c|c|c|c|}\toprule 

\multicolumn{2}{|l|}{${\bf Particles}$}&\multicolumn{10}{c|}{\bf Multiplicity class} \\
\cline{3-12}
\multicolumn{2}{|c|}{} &{\bf Mul1} & {\bf Mul2} & {\bf Mul3} & {\bf Mul4}& {\bf Mul5} &{\bf Mul6} & {\bf Mul7} & {\bf Mul8} & {\bf Mul9}&{\bf Mul10}\\
\hline

%%%%%%%%%%%%%%%%%%%%%%%%%%%%%%% -----   PI ------   %%%%%%%%%%%%%%%%%%%%%%%%%%%%%%%%%%%%%%%%%%%
\multirow{3}{*}{$\bf{\pi^{\pm}}$}&T (GeV)&0.093 $\pm$ 0.001 &0.089 $\pm$ 0.001 &0.087 $\pm$ 0.001 &0.085 $\pm$ 0.001 &0.083 $\pm$ 0.001 &0.081 $\pm$ 0.001 &0.078 $\pm$ 0.001 &0.076 $\pm$ 0.001 &0.073 $\pm$ 0.001 &0.068 $\pm$ 0.001\\
                   
                   \cline{2-12} 
                    & q &1.163 $\pm$ 0.001 &1.162 $\pm$ 0.001 &1.161 $\pm$ 0.001 &1.161 $\pm$ 0.001 &1.160 $\pm$ 0.001 &1.159 $\pm$ 0.001 &1.157 $\pm$ 0.001 &1.156 $\pm$ 0.001 &1.152 $\pm$ 0.001 &1.143 $\pm$ 0.001\\
                    \cline{2-12}
                    & $\chi^2$/NDF&7.574  &7.757  &6.792  &5.950 &5.294  &4.349  &3.074  &1.929  &0.603  &0.663\\
\hline
%%%%%%%%%%%%%%%%%%%%%%%%%%%%%%%%%%%%%%%%%%%%%%%%%%%%%%%%%%%%%%%%%%%%%%%%%%%%
%%%%%%%%%%%%%% KK
%%%%%%%%%%%%%%%%%%%%%%%%%%%%%%%%%%%%%%%%%%%%%%%%%%%%%%%%%%%%%%%%%%%%%%%%%%
\multirow{3}{*}{$\bf{K^{\pm}}$}&T (GeV)&0.148 $\pm$ 0.003 &0.135 $\pm$ 0.003 &0.125 $\pm$ 0.003 &0.118 $\pm$ 0.003 &0.113 $\pm$ 0.003 &0.107 $\pm$ 0.003 &0.098 $\pm$ 0.003 &0.090 $\pm$ 0.003 &0.080 $\pm$ 0.002 &0.060 $\pm$ 0.002 \\
                   
                   \cline{2-12} 
                    & q &1.143 $\pm$ 0.002 &1.146 $\pm$ 0.002 &1.148 $\pm$ 0.002 &1.149 $\pm$ 0.002 &1.150 $\pm$ 0.002 &1.151 $\pm$ 0.002 &1.152 $\pm$ 0.002 &1.153 $\pm$ 0.002 &1.153 $\pm$ 0.002 &1.150 $\pm$ 0.002\\
                    \cline{2-12}
                    & $\chi^2$/NDF&0.290  &0.236  &0.255  & 0.200 &0.155  &0.130  &0.131  &0.084  &0.062  &0.145\\

%%%%%%%%%%%%%%%%%%%%%%%%%%%%%%%%%%%%%%%%%%%%%%%%%%%%%%%%%%%%%%%%%%%%%%%%%%%%

%%%%%%%%%%%%% PPbar
%%%%%%%%%%%%%%%%%%%%%%%%%%%%%%%%%%%%%%%%%%%%%%%%%%%%%%%%%%%%%%%%%%%%%%%%%%
\multirow{3}{*}{$\bf{p + \overline{p}}$}&T (GeV)&0.183 $\pm$ 0.006 &0.160 $\pm$ 0.005 &0.143 $\pm$ 0.005 &0.131 $\pm$ 0.005 &0.119 $\pm$ 0.005 &0.101 $\pm$ 0.004 &0.090 $\pm$ 0.004 &0.075 $\pm$ 0.004 &0.049 $\pm$ 0.004 &0.017 $\pm$ 0.001\\
                   
                   \cline{2-12} 
                    & q &1.112 $\pm$ 0.003 &1.115 $\pm$ 0.002 &1.117 $\pm$ 0.002 &1.118 $\pm$ 0.002 &1.121 $\pm$ 0.002 &1.126 $\pm$ 0.002 &1.126 $\pm$ 0.002 &1.127 $\pm$ 0.002 &1.135 $\pm$ 0.002 &1.136 $\pm$ 0.001\\
                    \cline{2-12}
                    & $\chi^2$/NDF&0.483  &0.701  &0.622  & 0.384 &0.554  &0.637  &0.452  &0.297  &0.501  &0.331\\
\hline
%%%%%%%%%%%%%%%%%%%%%%%%%%%%%%%%%%%%%%%%%%%%%%%%%%%%%%%%%%%%%%%%%%%%%%%%%%%%

\multirow{3}{*}{$\bf{K^{0}_{S}}$}& T (GeV) &0.152$\pm$0.001&0.137$\pm$0.001&0.131$\pm$0.004&0.124$\pm$0.003&0.119$\pm$0.005&0.111$\pm$0.004&0.103$\pm$0.004&0.095$\pm$0.002&0.085$\pm$0.003&0.068$\pm$0.003\\
\cline{2-12}
                   
                    & q &1.141$\pm$0.001 & 1.144$\pm$0.001 & 1.144$\pm$0.002&1.145$\pm$0.002 &1.146$\pm$0.003&1.148$\pm$0.002&1.148$\pm$0.002&1.150$\pm$0.002&1.150$\pm$0.002&1.147$\pm$0.002\\
                    \cline{2-12}
                    & $\chi^2$/NDF& 0.275&0.429&0.235&0.323&0.309&0.286&0.484&0.384&0.321&0.494\\  
\hline
%%%%%%%%%%%%%%%%22222222222%%%%%%%%%%%%%%%%%%%%%%%%%
\multirow{3}{*}{$\bf{\Lambda+\bar{\Lambda}}$}&T (GeV)&0.245$\pm$ 0.0&0.201$\pm$ 0.0&0.179$\pm$0.0&0.159$\pm$0.0&0.146$\pm$0.0&0.128$\pm$0.0&0.102$\pm$0.0&0.082$\pm$0.0&0.056$\pm$0.0&0.010$\pm$0.0\\
                   
                   \cline{2-12} 
                    & q & 1.086$\pm$0.006& 1.097$\pm$0.004& 1.101$\pm$0.007&1.106$\pm$0.004 &1.108$\pm$0.004&1.111$\pm$0.001&1.118$\pm$0.003&1.123$\pm$0.002&1.128$\pm$0.004&1.139$\pm$.001\\
                    \cline{2-12}
                    & $\chi^2$/NDF& 0.554&0.543&0.281&0.311&0.272&0.307&0.201&0.160&0.312&0.248\\

\hline
%%%%%%%%%%%%%%%%333333333333%%%%%%%%%%%%%%%%%%%%%%%%%
\multirow{3}{*}{$\bf {\Xi^{-}+\bar{\Xi}^{+}}$}&T (GeV)&0.308$\pm$0.0&0.260$\pm$0.0&0.224$\pm$0.0&0.212$\pm$0.0&0.186$\pm$0.0&0.164$\pm$0.0&0.147$\pm$0.0&0.122$\pm$0.0&0.074$\pm$0.0 &0.045$\pm$0.001\\
                   
                   \cline{2-12} 
                    & q &1.069$\pm$0.015 & 1.081$\pm$0.005&1.086$\pm$0.005 &1.088$\pm$0.004 &1.096$\pm$0.003&1.100$\pm$0.003&1.101$\pm$0.003&1.108$\pm$0.002&1.121$\pm$0.002&1.122$\pm$0.002\\
                  \cline{2-12}
                    & $\chi^2$/NDF& 0.837&0.458&0.350&0.133&0.168&0.232&0.237&0.543&0.313&0.369\\

 \hline 
 
 \multicolumn{2}{|c|}{} &{\bf Mul1} & {\bf Mul2} & {\bf Mul3} & \multicolumn{2}{|c|}{\bf Mul[4 + 5]} &{\bf Mul6} & {\bf Mul7} & {\bf Mul8} & {\bf Mul9}&{\bf Mul10}\\
\hline
                    
                %%%%%% % %phi % % % % % %% %

\multirow{3}{*}{$\bf{\Phi}$}&T (GeV)&0.241 $\pm$	0.027&0.212 $\pm$ 0.018&0.181 $\pm$ 0.017&\multicolumn{2}{|c|} {0.177 $\pm$ 0.015} &0.161 $\pm$ 0.015&0.116$\pm$ 0.015&0.109 $\pm$ 0.018&0.070 $\pm$ 0.018&0.019 $\pm$ 0.002\\
                   
                   \cline{2-12} 
                    & q &1.116 $\pm$ 0.012 &1.122 $\pm$ 0.008 &1.131 $\pm$ 0.008 &\multicolumn{2}{|c|} {1.127 $\pm$ 0.006 }&1.128 $\pm$ 0.007 &1.145 $\pm$ 0.007 &1.141 $\pm$ 0.009 &1.153 $\pm$ 0.009 &1.159 $\pm$ 0.002\\
                    \cline{2-12}
                    & $\chi^2$/NDF&1.790  &0.687  &0.840  & \multicolumn{2}{|c|} {0.268  } &0.666  &0.293  &0.479  &0.479  &0.159\\
% % % % % % % % % % % % % % % % % % % % % % % % %%%%%%%%%%%%%%%%%%%%%%%%%%%%%%%%

                %%%%%% % %KStar % % % % % %% %
\hline

\multirow{3}{*}{$\bf{K^{*0} + \overline {K^{*0} }}$}&T (GeV)&0.225 $\pm$ 0.029 &0.230 $\pm$ 0.021 &0.185 $\pm$ 0.020 &\multicolumn{2}{|c|} {0.172 $\pm$ 0.018 } &0.132 $\pm$ 0.017 &0.124 $\pm$ 0.019 &0.109 $\pm$ 0.019 &0.076 $\pm$ 0.017 &0.024 $\pm$ 0.019\\
                   
                   \cline{2-12} 
                    & q &1.117 $\pm$ 0.013 &1.112 $\pm$ 0.009 &1.126 $\pm$ 0.009 &\multicolumn{2}{|c|} {1.126 $\pm$ 0.009  }&1.137 $\pm$ 0.008 &1.134 $\pm$ 0.010 &1.135 $\pm$ 0.010 &1.145 $\pm$ 0.009 &1.155 $\pm$ 0.011\\
                    \cline{2-12}
                    & $\chi^2$/NDF&0.394  &0.615  &0.806  &\multicolumn{2}{|c|} {0.681   } &0.587  &0.807  &0.339  &0.165  &0.268\\
                    
                    \hline

\multicolumn{2}{c|}{}&\multicolumn{2}{c|}{\bf Mul[1+2]}&\multicolumn{2}{c|}{\bf Mul[3+4]}&\multicolumn{2}{c|}{\bf Mul[5+6]}&\multicolumn{2}{c|}{\bf Mul[7+8]}&\multicolumn{2}{c|}{\bf Mul[9+10]} \\
\hline
\multirow{3}{*}{$\bf{\Omega^{-}+\bar{\Omega}^{+}}$}& T (GeV) &\multicolumn{2}{c|}{0.422$\pm$0.0}&\multicolumn{2}{c|}{0.310$\pm$0.0}&\multicolumn{2}{c|}{0.330$\pm$0.001}&\multicolumn{2}{c|}{0.104$\pm$0.0}&\multicolumn{2}{c|}{0.052$\pm$0.0}\\
 \cline{2-12} 
  & q &\multicolumn{2}{c|}{1.035$\pm$0.028} & \multicolumn{2}{c|}{1.066$\pm$0.011}&\multicolumn{2}{c|}{1.044$\pm$0.005}&\multicolumn{2}{c|}{1.117$\pm$0.015}&\multicolumn{2}{c|}{1.124$\pm$0.006} \\
 \cline{2-12}
 & $\chi^2$/NDF& \multicolumn{2}{c|}{0.297}&\multicolumn{2}{c|}{0.478}&\multicolumn{2}{c|}{0.092}&\multicolumn{2}{c|}{0.238}&\multicolumn{2}{c|}{0.652}\\
 \hline 
 \end{tabular}

%\end{sidewaystable}
\end{adjustbox}
% \end{center}
 \end{table*}
%\end{widetext}
%\clearpage

%%%%%%%%%%%%%%%%%%%%%%%%%%%%%%%%%%%%%%%%%%%%%%%%%%%%%%%%

\bef[H]
\begin{center}
\includegraphics[scale=0.44]{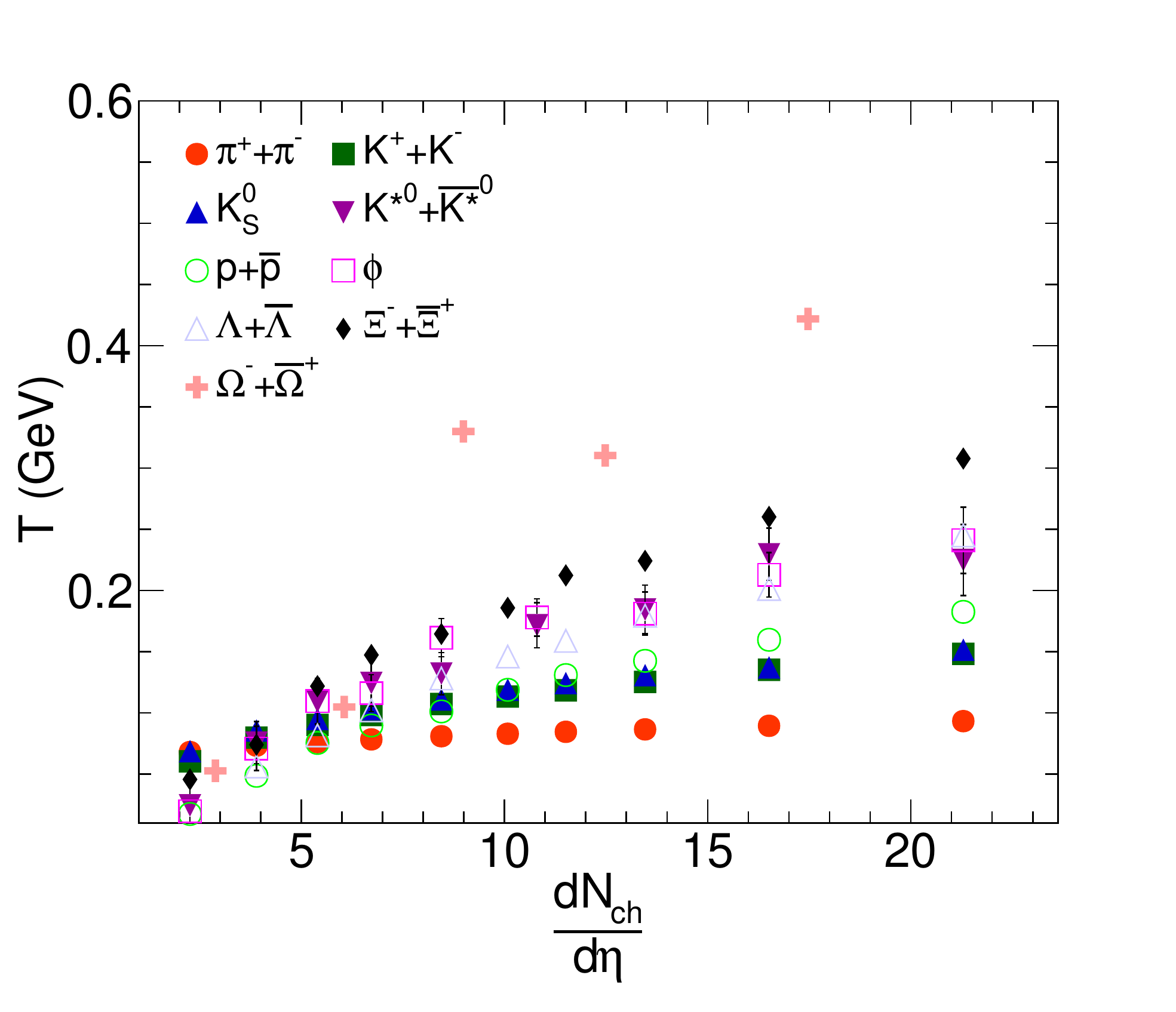}
\caption{(color online) Multiplicity dependence of $T$ for $pp$-collisions at $\sqrt{s}$ = 7 TeV using Eq.~\ref{eq6} as a fitting function. }
\label{fit:Tsallis:T}
\end{center}
\eef

\bef[H]
\begin{center}
\includegraphics[scale=0.44]{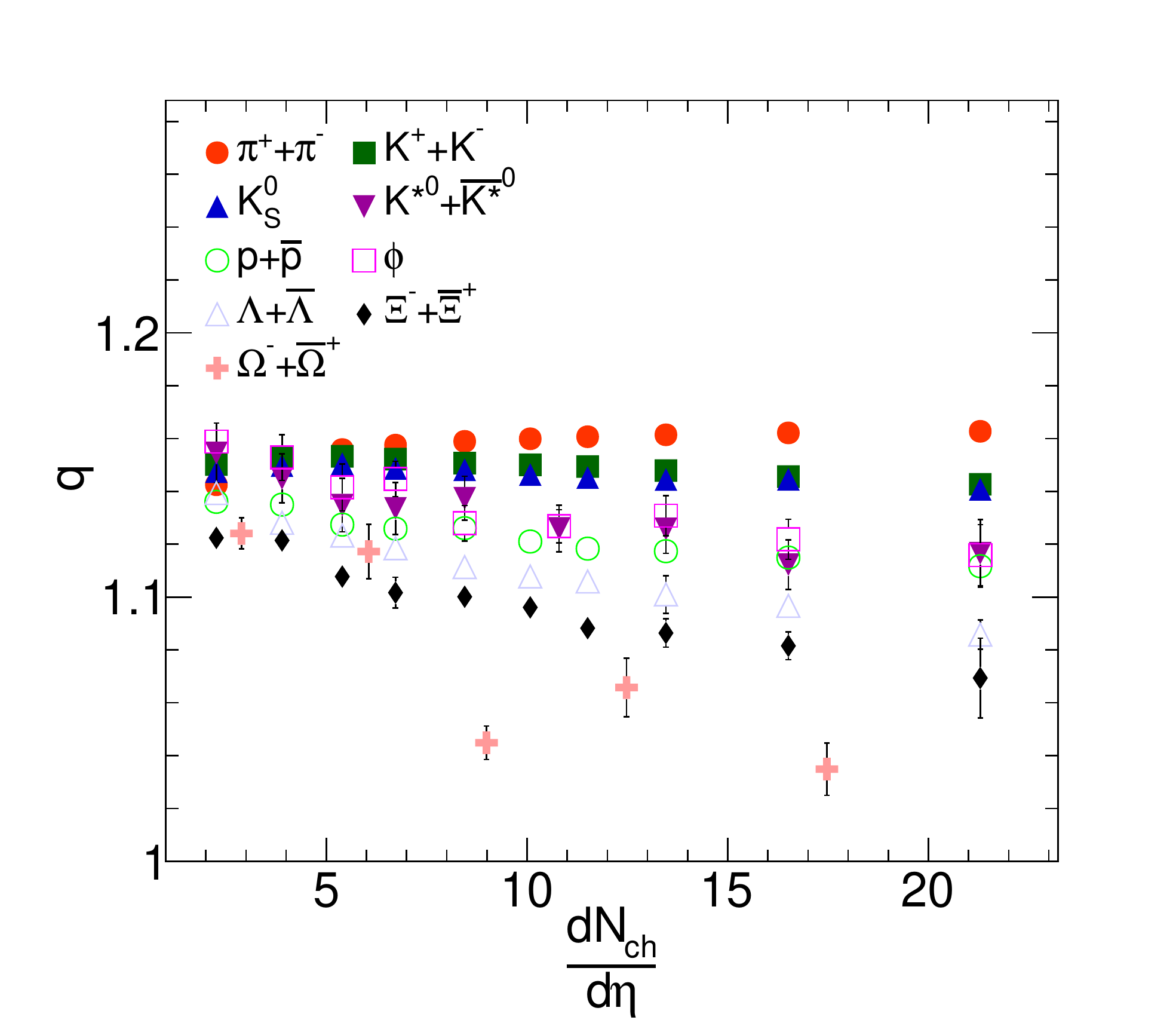}
\newline
\caption{(color online) Multiplicity dependence of the non-extensive parameter, $q$  for $pp$-collisions at $\sqrt{s}$ = 7 TeV using Eq.~\ref{eq6} as a fitting function.}
\label{fit:Tsallis:q}
\end{center}
\eef

\section{Summary}
\label{sec3}
Recently, high-multiplicity events in $pp$-collisions at the LHC energies have drawn considerable interest to the research community, as it has shown heavy-ion like properties {\it e.g.}, enhanced production of strange particles \cite{Adam:2016emw}, which are yet to be understood. In this work, we have tried to understand these events from the perspective of thermodynamics. The information regarding kinetic freeze-out temperature of the identified particles at this energy is estimated by fitting BGBW function upto low-$p_{\rm T}$. To address the high-$p_{\rm T}$, which has  pQCD inspired power-law contribution become customary to use a thermodynamically consistent Tsallis non-extensive statistics to describe the complete spectra.

We have analysed the  multiplicity dependence of the $p_{\rm T}$-spectra of identified hadrons in $pp$-collisions at $\sqrt{s}$ = 7 TeV measured by the ALICE experiment at the LHC, using BGBW model and thermodynamically consistent non-extensive statistics. The extracted thermodynamic parameters $i.e.$ the kinetic freeze-out parameter ($T_{\rm kin}$) and radial flow ($\beta$) are studied as a function of charged particle multiplicity in BGBW formalism. Similarly, we have studied the Tsallis temperature parameter ($T$) and the non-extensive parameter ($q$) as a function of charged particle multiplicity using Tsallis statistics. In addition to this, we have also studied these parameters as a function of particle mass.
In summary,

\begin{itemize}

\item It is observed that BGBW model explains the experimental data upto $p_{\rm T}$ $\simeq$ 3 GeV/$c$ with an appreciable $\chi^2$/NDF.   The multistrange particles are not included in the fitting due to large statistical uncertainties.    

\item We have extracted the kinetic freeze-out temperature for all the identified hadrons. It is discovered that $T_{\rm kin}$ follows a mass dependent pattern and acquires higher values for heavier particles. This goes inline with the fact that heavier particles freeze-out earlier in time. However, we notice a multiplicity independent behaviour of $T_{\rm kin}$ particularly for lighter hadrons.    

\item The radial flow ($\beta$) parameter is also extracted in this study, which is observed higher for lighter particles. This observation reveals hydrodynamic behaviour of particles. 

\item The near-multiplicity independent behaviour of radial flow velocity, $\beta$ is an important observation and this needs further investigations.

\item It has been manifested in the present paper, the Tsallis distribution provides a complete description of identified particle spectra produced in $pp$-collisions at $\sqrt{s}$ = 7 TeV upto very high-$p_{\rm T}$.\\

\item The variable $T$ shows a systematic increase with multiplicity, the heaviest baryons showing the steepest increase. This is an indication of a mass hierarchy in particle freeze-out.\\

\item The obtained parameters show variations with the event multiplicity. The notable variation of the non-extensive parameter, $q$ which decreases towards the value 1 as the multiplicity increases and this effect is more significant for heavy mass particles. This shows the tendency of the produced system to equilibrate with higher multiplicities. This goes inline with the expected multi-partonic interactions, which increase for higher multiplicities in $pp$-collisions and are thus responsible for bringing the system towards thermodynamic equilibrium~\cite{Thakur:2017kpv}.
\end{itemize}

Conclusively, we find that BWBG explains the transverse momentum spectra up to $p_{\rm T} \simeq$ 3 GeV/$c$ for high-multiplicity $pp$-collisions with an appreciable $\chi^2$/NDF, while the Tsallis statistics describe the complete spectra for all the event classes. 

\section*{Acknowledgements}
The authors acknowledge the financial supports from ALICE Project No. SR/MF/PS-01/2014-IITI(G) of Department of Science \& Technology, Government of India.

\end{document}